\documentstyle[aps,manuscript,epsfig]{revtex}

\begin{document}

\draft

\title{The Halo of $^{14}$Be}

\author{
M.~Labiche$^a$\cite{now1}, N.A.~Orr$^a$, F.M.~Marqu\'{e}s$^a$, 
J.C.~Ang\'{e}lique$^a$, L.~Axelsson$^b$, 
B.~Benoit$^c$, U.C.~Bergmann$^d$, M.J.G.~Borge$^e$, W.N.~Catford$^f$, 
S.P.G.~Chappell$^g$,  
N.M.~Clarke$^h$, G.~Costa$^i$, N.~Curtis$^f$\cite{now2}, A.~D'Arrigo$^c$, 
E.~de~G\'{o}es~Brennand$^c$, O.~Dorvaux$^i$, G.~Fazio$^j$, M.~Freer$^{a,h}$,  
B.R.~Fulton$^h$,  G.~Giardina$^j$, 
S.~Gr\'{e}vy$^{k}$\cite{now3}, 
D.~Guillemaud-Mueller$^{k}$, F.~Hanappe$^c$, B.~Heusch$^i$, K.L.~Jones$^f$, 
B.~Jonson$^b$,
C.~Le~Brun$^a$, S.~Leenhardt$^k$, M.~Lewitowicz$^l$, M.J.~Lopez$^l$,
K.~Markenroth$^b$, A.C.~Mueller$^k$, T.~Nilsson$^b$\cite{now6}, A.~Ninane$^{a}$\cite{now4}, G. Nyman$^b$,
F.~de~Oliveira$^l$, I.~Piqueras$^e$, K.~Riisager$^d$,
M.G.~Saint~Laurent$^l$, F.~Sarazin$^l$\cite{now5}, S.M.~Singer$^h$, O.~Sorlin$^k$, 
L.~Stuttg\'{e}$^i$
}

\address{ 
$^a$ Laboratoire de Physique
Corpusculaire, ISMRA et Universit\'{e} de Caen,  IN2P3-CNRS, 14050 Caen 
Cedex, France.\\ 
$^b$Experimentall Fysik, Chalmers Tekniska H\"{o}gskola,
S-412 96 G\"{o}teborg, Sweden.\\ 
$^c$Universit\'{e} Libre de Bruxelles, PNTPM, CP 226,
B-1050 Bruxelles, Belgium.\\ 
$^d$Det Fysiske Institut, Aarhus Universitet, DK
8000 Aarhus C, Denmark.\\ 
$^e$Instituto de Estructura de la Materia, CSIC, E-28006 Madrid, Spain.\\ 
$^f$Department of Physics, University of Surrey,
Guildford, GU2~7XH, U.~K.\\ 
$^g$Nuclear and Astrophysics Laboratory,
University of Oxford, Oxford OX1 3RH, U.~K.\\ 
$^h$School of Physics and Astronomy, University of Birmingham, 
Birmingham~B15~2TT, U.~K.\\ 
$^i$Institut de
Recherches Subatomique, IN2P3-CNRS, Universit\'{e} Louis Pasteur, 67037
Strasbourg Cedex, France.\\ 
$^j$Dipartimento di Fisica, Universit\`{a} di Messina, I-98166 Messina,
Italy. \\ 
$^k$Institut de Physique Nucl\'{e}aire, IN2P3-CNRS, 91406 Orsay Cedex,
France. \\ 
$^l$GANIL (CEA/DSM-CNRS/IN2P3), BP 5027, 14076
Caen Cedex, France.\\ 
} 

\date{Received \today}

\maketitle

\newpage

\begin{abstract}

The two-neutron halo nucleus $^{14}$Be has been investigated in a
kinematically complete measurement of the fragments ($^{12}$Be and neutrons) 
produced in dissociation at 35 MeV/nucleon on C and Pb
targets.  Two-neutron removal cross-sections, neutron angular
distributions and invariant mass spectra characteristic of a halo
were observed and the electromagnetic (EMD) contributions deduced. Comparison 
with three-body model predictions indicate that the halo
wavefunction contains a large $\nu (2s_{1/2})^{2}$ admixture. 
The EMD invariant
mass spectrum exhibited a relatively narrow
structure near threshold
(E$_{decay}$=1.8$\pm$0.1 MeV, $\Gamma$ = 0.8$\pm$0.4 MeV) consistent with
a soft-dipole excitation.

\end{abstract}

\bigskip

\pacs{PACS number(s): 27.20.+n6, 25.60.Dz, 25.60.Ge, 24.30.Gd }

The size and distribution of matter in the nucleus have long played a central
role in nuclear physics.  Indeed, such gross properties reflect the combined
effects of many fundamental aspects of nuclei. For the
stable nuclei, measurements employing conventional probes, such as high energy
electron and hadron scattering, have shown that the neutron and proton
distributions exhibit essentially identical radii \cite{Bat89}. In contrast,
for some light nuclei far from stability, which combine a large neutron
excess with very weak binding, large differences have been found. 
Such ``halo'' systems are well described by a core, resembling a normal nucleus, 
surrounded by an extended valence neutron density distribution \cite{Han95}. 

In general terms the halo may be regarded as a threshold phenomenon whereby the 
loosely bound valence neutrons tunnel with significant 
probability into the classically forbidden region outside the core potential.  
Within a simple 
quasideuteron description, the extent of the halo is governed 
by the separation energy and reduced mass of the system \cite{Han87}.  
Under more realistic 
considerations the development of the halo is also influenced by the 
centrifugal barrier \cite{Rii92Fed93}.
In the cases of $^{6,8}$He, $^{11}$Be and $^{11}$Li, which have been 
investigated 
experimentally in 
considerable detail, the valence neutrons occupy the $2s_{1/2}$ and/or 
$1p_{3/2,1/2}$
single-particle orbitals.  In $^{14}$Be
the configuration of the halo neutrons would, in a na\"ive shell model 
prescription,
be $\nu(d_{5/2})^2$.  Sophisticated models suggest, however, that a 
$\nu(s_{1/2})^2$ admixture is also present \cite{Ren95,Ada95,Tho96,Bay97}.  
Unfortunately, a paucity 
of experimental data 
\cite{Rii92a,Zah93,Zah94}
has precluded the elucidation of the structure of $^{14}$Be beyond the 
matter radius \cite{Tan88,Lia90,AlK96,Suz99}.
Compared to the other halo systems, the comparatively strong binding 
of the valence neutrons in $^{14}$Be
(S$_{2n}$=1.34$\pm$0.11MeV \cite{Gil84,Wou88}) combined with the $\nu(d_{5/2})^2$
component may 
provide a new window on continuum excitations,
including the long sought-after Soft-Dipole Resonance (SDR) \cite{Ike92,Kob89}.

The goal of the present study was thus to explore the halo structure and 
continuum
excitations of the two-neutron halo nucleus $^{14}$Be.  The tool chosen was a 
kinematically
complete measurement of the fragments ($^{12}$Be and two neutrons) from the
dissociation of an intermediate energy beam of $^{14}$Be on C and Pb targets.
Such a measurement allowed the two-neutron removal cross sections,
neutron angular distributions and invariant mass spectra to be
extracted (the results of an analysis of the neutron-neutron correlations
have been presented elsewhere \cite{Mar00a}).  The use of C and Pb 
targets permitted the electromagnetic
component of the dissociation (EMD) to be deduced.

The $^{14}$Be beam ($\sim$130 pps) was prepared using the 
LISE3 spectrometer and a 63 MeV/nucleon $^{18}$O primary beam bombarding a
thick Be production target.  The mean energy of the 
beam at the mid-point of the secondary breakup targets was 35 MeV/nucleon. 
The energy spread in the beam was 10\% 
and was
compensated for by a time-of-flight (TOF) measurement over a 24~m
flight-path between a parallel-plate avalanche counter (PPAC) located at 
the first 
focus of the spectrometer and the beam identification Si-detector.
The beam particles were tracked onto the breakup targets 
(C 275 mg/cm$^2$, Pb 570 mg/cm$^2$) using two position
sensitive PPAC's (resolution FWHM $\approx$ 1-2 mm).  
Owing to the mixed nature of the secondary beam
(50 \% $^{14}$Be) the incoming ions were identified on a particle-by-particle
basis using the TOF information combined with the energy loss derived
from a Si-detector (300 $\mu$m) located just upstream of the target.
The charged fragments from breakup were identified using a large area
(5$\times$5~cm$^2$) position sensitive
(FWHM $\approx$ 0.5mm) 
Si-CsI telescope (Si 500 $\mu$m, CsI 2.5 cm) centred at zero degrees and
located 11.4 cm
downstream of the target.  The 
energy response of the telescope (FWHM = 1.5\%) was calibrated using 
various mixed secondary
beams containing $^{12}$Be with energies straddling that expected for
$^{12}$Be fragments arising from the dissociation of $^{14}$Be.  
In order to account for events arising from reactions in the telescope, 
data was also acquired without a reaction target with the beam energy reduced by 
the
amount corresponding to the energy loss in the C and Pb targets.   

The neutrons emitted at forward angles were detected using the 99 elements of
the D\'eMoN array \cite{Mar00b}.  The array covered angles between 
+13$^\circ$ and -40$^\circ$ in the horizontal plane and $\pm$14$^\circ$ in the 
vertical
with the modules arranged in a staggered configuration at distances between 2.5 
and 6.5 m 
from the target \cite{Mar00b}.  Such a geometry provided for  
a relatively high two-neutron detection efficiency (1.5\%) 
whilst reducing the rate of cross-talk --- both intrinsically and
via the use of an off-line rejection algorithm --- 
to negligible levels \cite{Mar00b,Lab99}.
A threshold of 15 MeV on the neutron energy was applied in the off-line analysis to
eliminate contamination from the small number of evaporation neutrons arising 
from the target.

The results obtained for the two-neutron removal cross sections, 
$\sigma_{-2n}$
($^{12}$Be identified in the telescope), the single-neutron 
angular distributions,
$d\sigma/d\Omega$ ($^{12}$Be and neutron), and the associated 
angle integrated 
(0-40$^\circ$) 
cross sections, $\sigma_n$, are displayed in table I and 
figure 1a. 
In addition, the average neutron multiplicities have been derived 
($\overline{m}_n = \sigma_n/\sigma_{-2n}$) and are also listed.  
The single-neutron angular distributions are well characterised by 
a Lorentzian lineshape \cite{Rii92a,Lab99} and the corresponding momentum
width parameters, $\Gamma_n$, have been tabulated.
The large neutron removal cross sections and relatively narrow neutron 
distributions, while not as pronounced as for $^{11}$Li \cite{Rii92a}, 
clearly 
indicate the halo character of $^{14}$Be.  The present results improve 
considerably 
on the 
earlier measurements of Riisager {\em et al.} \cite {Rii92a} 
which suffered from poor
statistics (no angular distribution could be constructed for a heavy target) and
were restricted to a limited angular range.

The multiplicities obtained for the two targets are instructive in terms of
the reaction mechanisms leading to dissociation \cite{Bar93}.  
For a light target, unless the halo neutrons are highly spatially 
correlated, 
the reaction is 
expected to proceed via single-neutron removal (absorption or diffraction) 
followed by
the in-flight decay of $^{13}$Be.  As approximately equal contributions 
are expected for 
absorption and diffraction \cite{Bar93} the average neutron multiplicity should be 1.5, 
in accordance with that measured here (table I).
This scenario is also supported by the single-neutron angular distribution 
for the C target which is 
well reproduced assuming passage via a
low-lying resonance in $^{13}$Be \cite{Lab99,Bar96}.
In the case of a heavy target, nuclear and Coulomb dissociation are
present.  Given that Coulomb dissociation should be associated with a 
multiplicity of 2,
the average multiplicity for dissociation on Pb should be between 1.5 and 2, 
as observed.

The enhanced cross section for dissociation on the Pb target
is indicative of a large EMD contribution.
Assuming that the nuclear--Coulomb interference is small, the C target data 
(which arises essentially from nuclear induced reactions) may
be scaled to estimate the nuclear contribution to breakup on Pb \cite{Kob89,Lab99}. 
Assuming a root-mean-square  
radius of 3.2 fm for $^{14}$Be \cite{AlK96,Suz99}, 
$\sigma_{-2n}^{nucl}(Pb)$ = 0.85$\pm$0.07~b
and, consequently,  $\sigma_{-2n}^{EMD}(Pb)$ = 1.45$\pm$0.40~b.
The latter can be compared to the value of 0.47$\pm$0.15~b measured at 800 MeV/nucleon 
\cite{Kob89}.  Importantly, for halo nuclei, the EMD cross section is dominated 
by the E1 component \cite{Ber92,Ale00}.
An enhancement with decreasing beam energy is thus 
expected,
owing to the large amount of dipole strength near threshold (see below) coupled
with the weighting of the virtual photon spectrum to low photon energies \cite{Hus91}.  

Assuming that the neutron angular distribution 
arising from nuclear dissociation on Pb
is identical to that measured for the C target, the single-neutron angular
distribution for EMD has been constructed (figure 1b) and the corresponding
integrated cross section and average multiplicity derived (table I).  
Interestingly, the angular distribution remains narrow and forward peaked
whilst the multiplicity is consistent 
with the value of 2 expected for EMD, confirming the validity of the
methods used to estimate the contribution arising from nuclear breakup.

The invariant mass spectra, reconstructed from the measured momenta of the beam
and fragments ($^{12}$Be and two neutrons) from breakup, are displayed in 
figure 2a and b for the C and Pb
targets.  The EMD spectrum (figure 2c) has
been deduced, as described above, following subtraction of the estimated 
nuclear contribution to 
reactions on Pb.  As for the spectra obtained with the C and Pb targets, the 
EMD spectrum exhibits enhanced strength around 2 MeV decay energy ($E_{decay}$).
Given the complex nature of the response function of the present setup, a
detailed Monte Carlo simulation, including the influence of all 
nonactive materials,
was developed based on the GEANT package \cite{Lab99}. 
The results shown in figure 2 were obtained following the descriptions for
dissociation on C and Pb outlined earlier.  In the case of the nuclear
induced reactions a single low-lying state 
in $^{13}$Be ($E_0 = 0.5$~MeV, $\Gamma_0 = 0.5\pm0.4$~MeV) was assumed 
to be populated following the diffraction of one of the halo neutrons \cite{Lab99,Jon00}.
The EMD was simulated under 
the assumption that the energy sharing between the $^{12}$Be and the 
two neutrons was governed by
3-body phase space.  As shown in figure 2c, the observed EMD decay energy spectrum
could be reproduced using a 
Breit-Wigner lineshape with a resonance 
decay energy of 
$E_0 = 1.8\pm0.1$ MeV and width $\Gamma_0 = 0.8\pm0.4$ MeV.  Furthermore, the
corresponding simulations of the  
single-neutron angular distributions were in 
good agreement with
those observed for reactions on C and Pb, as well as that deduced for EMD \cite{Lab99}.

As noted above, the EMD of halo nuclei is essentially E1 in character.
An analytical estimate for the E1 strength for two-neutron halo 
nuclei
has been 
derived in a simple 3-body model based on Yukawa wavefunctions \cite{Pus96}, 
whereby the maximum occurs for 
$E_{decay} = 6/5S_{eff}$, where $S_{eff}\approx1.5S_{2n}$.  Whilst agreeing 
well with the
available results for $^{11}$Li, a maximum is predicted for $^{14}$Be at 
$E_{decay} \approx 2.4$~MeV, somewhat above that
observed here.  

Thompson and Zhukov have examined 
$^{14}$Be within the framework of a more realistic 3-body model 
in which the $^{12}$Be core is treated as inert  
\cite{Tho96} and a number of
trial wavefunctions developed.  Based on the binding energy and 
matter radius \cite{AlK96,Suz99} of $^{14}$Be, together with
the known d-wave resonance at 2.01~MeV in $^{13}$Be 
\cite{Ost92}, 
two $^{14}$Be wavefunctions are favoured (both of which
require an s-wave state near threshold in $^{13}$Be 
as suggested by recent experiments 
\cite{Jon00,Tho,Bel}): the so-called D4 
wavefunction
-- 86\% $\nu (2s_{1/2})^{2}$ and 10\% $\nu (1d_{5/2})^{2}$; and C7 --  
29\% $\nu (2s_{1/2})^{2}$ and 67\% $\nu (1d_{5/2})^{2}$.  The EMD decay energy 
spectra 
calculated for these wavefunctions for breakup at 35 MeV/nucleon on Pb 
\cite{Tho96} are compared in figure 3 with that of the
empirical Breit-Wigner deduced from the present measurements.  
The corresponding integrated two-neutron removal cross
sections are 1.05~b (D4) and 0.395~b (C7) \cite{Tho96}, compared to
the measured value of 1.45$\pm$0.40~b.  
Although the strength is predicted to be concentrated at a somewhat lower 
energy than that observed, a large 
$\nu (2s_{1/2})^{2}$ admixture to the valence neutrons wavefunction is favoured.
Such a result is supported by the total reaction cross section measurement of 
Suzuki {\em et al.} \cite{Suz99} and is also in line with Lagrange mesh calculations of 
the $^{14}$Be ground state (76\% $\nu (2s_{1/2})^{2}$, 18\% $\nu (1d_{5/2})^{2}$) 
\cite{Ada95,Bay97}.
It should be noted that the treatment of the core as 
inert precludes, ab initio, the existence of any simple 
negative parity resonances in $^{14}$Be.

Descouvemont has explored $^{13,14}$Be within a  
microscopic cluster ($^{12}$Be+n+n)
model in which the core is active \cite{Des95}. In the case of $^{13}$Be
an s-wave 
state is predicted very close 
to threshold, whilst the energy of the d-wave resonance is well reproduced.  
Significantly, a strong E1 transition [B(E1)$\approx$1.2e$^2$~fm$^2$] centred at 
$E_{decay}=1.5$~MeV
is predicted in $^{14}$Be, 
very close to the structure observed experimentally.  
Analysis of the corresponding energy surface suggests, however, that this
transition is not associated with a true resonance \cite{Des95}. 
Calculations of the form of the associated continuum energy spectrum would be
of considerable interest.
Further support for the predictions of this model exists in the observation 
in a heavy-ion double charge-exchange reaction
of a probable $2^+$ state in $^{14}$Be at $E_{decay} = 0.25\pm0.06$ MeV \cite{Bol95},
compared to a calculated value of 0.5 MeV.

It is interesting to note that
the width of the structure seen in the present experiment would correspond, in the case
of a true E1 resonance, to a  
mean lifetime ($1/\Gamma$) of some 250$\pm$120~fm/c.  This may 
be compared to a
simple $\hbar\omega$ ($E_x=S_{2n}+E_0=3.14\pm0.15$~MeV) collective mode 
oscillation period of $\sim$400~fm/c, suggesting again the nonresonant nature
of the observed transition.

In conclusion, the first kinematically complete breakup reaction study of
$^{14}$Be has been reported.  
Two-neutron removal cross sections, neutron angular distributions and invariant mass
spectra characteristic of a halo were measured.  
The EMD observables indicate that the configuration of the
halo neutrons contains a large $\nu (2s_{1/2})^{2}$ component.  
The relatively narrow structure observed near threshold in the EMD invariant mass 
spectrum is consistent with a soft-dipole excitation.  
Exploration of the continuum excitations beyond those probed here
($E_{decay}>$ 5~MeV) would thus be of particular interest. 
Additionally, spectroscopic studies of $^{13}$Be and 
a determination of the $\beta_2$ of $^{12}$Be,
which are essential to developing more refined models describing $^{14}$Be,
are needed.
Finally, in light of the present results, it would be highly desirable
to explore the $^{14}$Be continuum via other means, including inelastic  
scattering
and surface dominated probes such as transfer or charge exchange.

The authors are grateful to the support provided by the 
technical and operations staff of
LPC and GANIL.  Discussions with Ian Thompson and Pierre Descouvemont
are also acknowledged.  
This work was funded by the IN2P3-CNRS (France) and EPSRC (UK).  
Additional support was provided by the
ALLIANCE programme of the British Council and the Minist\`ere des
Affaires Etrang\`eres, and the Human Capital and Mobility Programme
of the European Community (Contract no. CHGE-CT94-0056).

\newpage

\begin{center}

FIGURE CAPTIONS\\

\end{center}

{\bf Figure 1:}  (a) Single-neutron angular distributions for dissociation on
C (open) and Pb (solid points).  The results for C have been scaled by a factor of 1.8
so as to represent the nuclear contribution to dissociation on Pb (see text).
(b) Deduced EMD single-neutron angular distribution for reactions on Pb.\\

{\bf Figure 2:}  Reconstructed $^{14}$Be decay energy spectra for dissociation on
(a) C, (b) Pb  and (c) that deduced for EMD on Pb.  The
histogrammes correspond to the results of simulations (see text).\\

{\bf Figure 3:}  Comparison of the EMD decay energy spectra for the 
3-body wavefunctions D4
and C7 \cite{Tho96} with that deduced from the present experiment -- $E_0=1.8$~MeV
and $\Gamma_0$=0.8~MeV (shaded region).  The later has been normalised to
an integrated cross section of 1.45$\pm$0.40b.\\

\begin{table}
\begin{center}
\begin{tabular}{ccccc} 
        & $\sigma_{-2n}$ [b] & $\sigma_n$ [b] & $\overline{m}_n$ & $\Gamma_n$ [MeV/c] \\
\hline 
C	& 0.46$\pm$0.04     & 0.75$\pm$0.10  & 1.6$\pm$0.3     & 75$\pm$3 \\
Pb      & 2.3$\pm$0.4       & 4.0$\pm$0.3    & 1.7$\pm$0.2     & 77$\pm$4 \\
Pb(EMD) & 1.45$\pm$0.40     & 2.7$\pm$0.4    & 1.9$\pm$0.6       & 87$\pm$6

\end{tabular}
\end{center}
\caption{Measured cross sections, average neutron multiplicities and neutron distribution 
momentum widths for the dissociation of $^{14}$Be at 35~MeV/nucleon.}
\end{table}

\newpage

\begin{figure}
\begin{center}
%\mbox{\psfig{file=prl_14be_fig1.eps,width=7.5cm}}
\mbox{\psfig{file=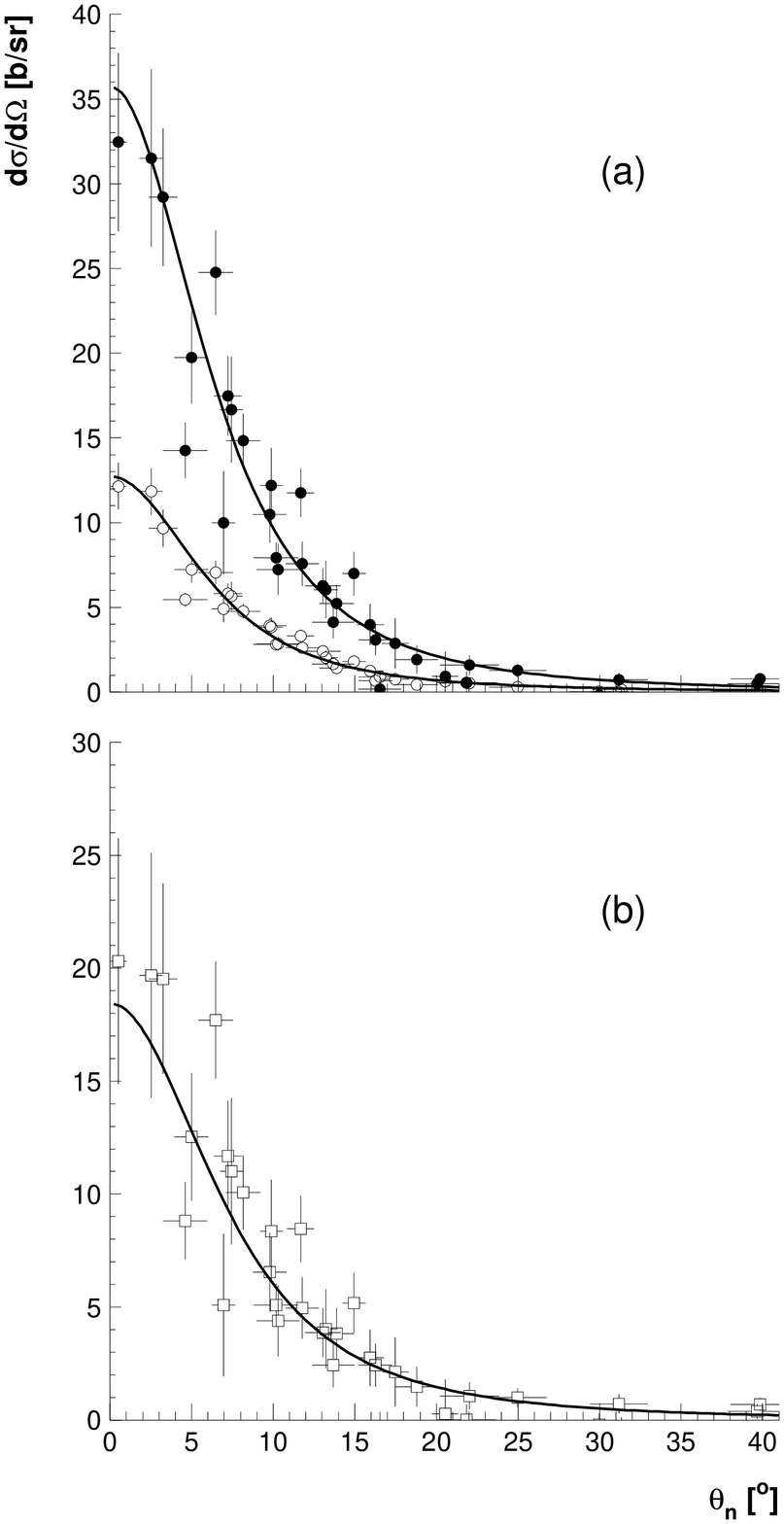,height=19cm}}
\end{center}
\end{figure}

\newpage

\begin{figure}
\begin{center}
%\mbox{\psfig{file=prl_14be_fig2.eps,width=7.5cm}}
\mbox{\psfig{file=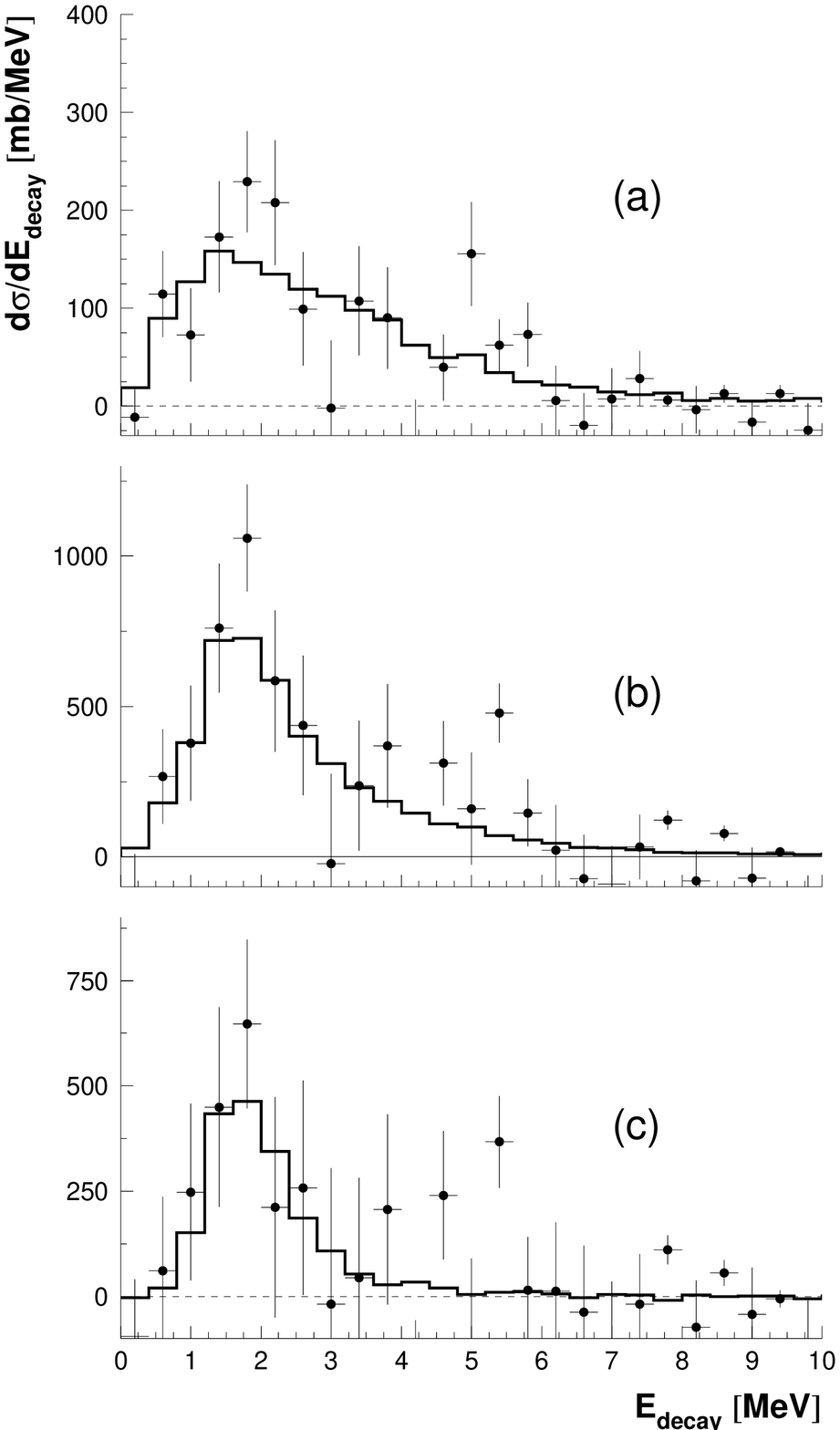,height=21cm}}
\end{center}
\end{figure}

\newpage

\begin{figure}
\begin{center}
%\mbox{\psfig{file=prl_14be_fig3.eps,width=7.5cm}}
\mbox{\psfig{file=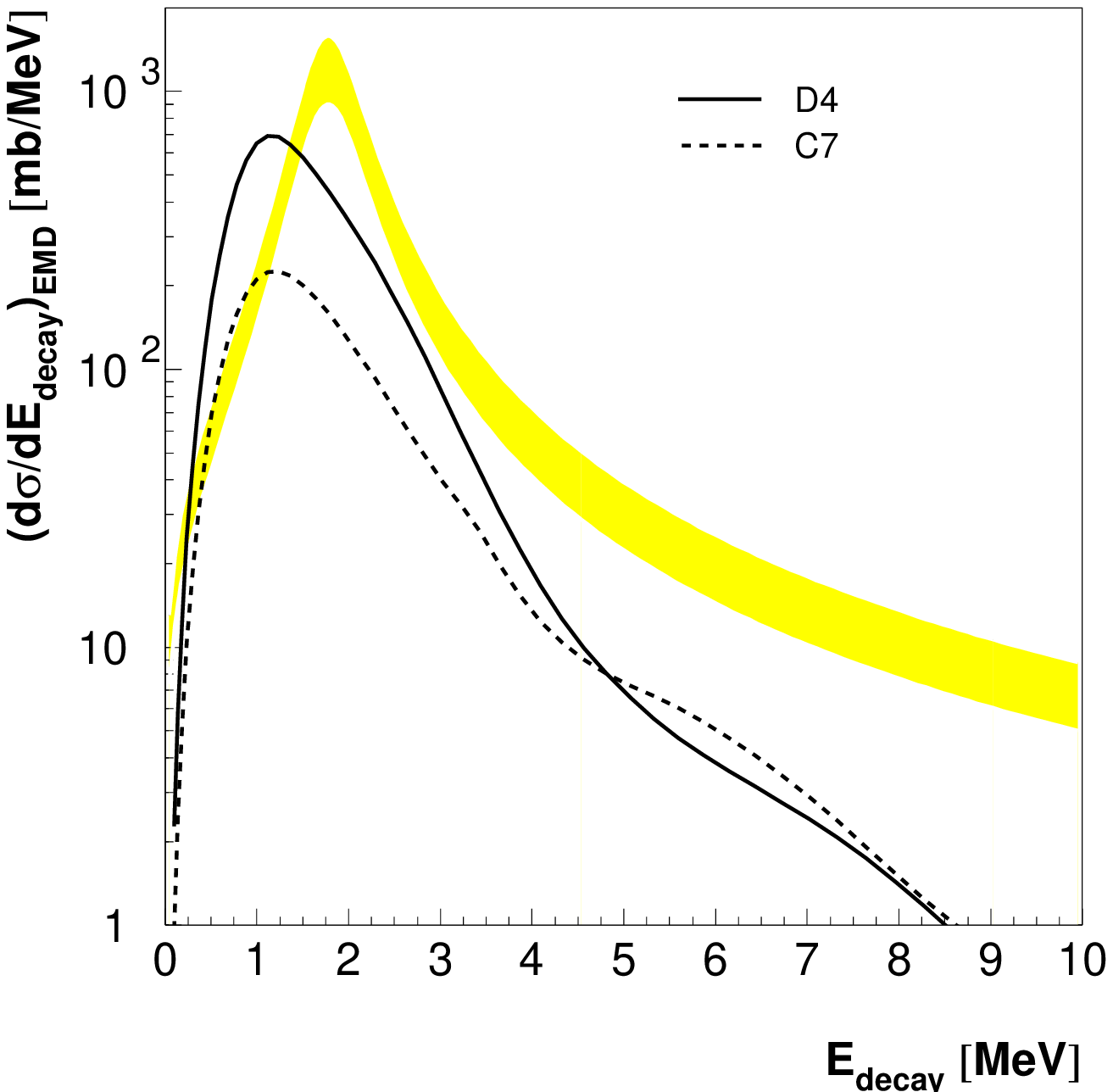,height=12cm}}
\end{center}
\end{figure}

\end{document}